\documentclass[journal,10pt]{IEEEtran}
\usepackage{graphicx}
\usepackage{epsfig}
\usepackage{epstopdf}
\usepackage{amsmath,amsthm,amsfonts,amssymb}
\usepackage{multirow}
\usepackage{cite}
\usepackage{tikz}
\usepackage{ellipsis}
\usepackage{xcolor}
\usepackage{xifthen}
\usepackage{cancel}
\usepackage[linesnumbered, ruled, vlined, noend]{algorithm2e}
\theoremstyle{definition} 
\theoremstyle{definition} 
\theoremstyle{definition} 
\theoremstyle{definition} 

\newcommand{\cblue}{\textcolor{blue}}
\newcommand{\cred}{\textcolor{red}}
\newcommand{\B}{\beta}
\newcommand{\Lm}{\lambda}
\newcommand{\fixme}[2]{\ifx&#2&{\leavevmode\color{red}#1}\else{\leavevmode\color{red}FIXME\{}#1{\leavevmode\color{red}\}}\footnote{{\leavevmode\color{red}#2}}\PackageWarning{Fixme}{#1: #2}\fi}
\pgfmathsetseed{1}
\usetikzlibrary{shapes,positioning,arrows,calc,graphs}
\usetikzlibrary{decorations.pathreplacing,decorations.markings,shapes.geometric}
\tikzset{naming/.style={align=center,font=\small}}
\tikzset{antenna/.style={insert path={-- coordinate (ant#1) ++(0,0.25) -- +(135:0.25) + (0,0) -- +(45:0.25)}}}
\tikzset{station/.style={naming,draw,shape=dart,shape border rotate=90, minimum width=10mm, minimum height=10mm,outer sep=0pt,inner sep=3pt}}
\tikzset{mobile/.style={naming,draw,shape=rectangle,minimum width=12mm,minimum height=6mm, outer sep=0pt,inner sep=3pt}}
\tikzset{radiation/.style={{decorate,decoration={expanding waves,angle=90,segment length=4pt}}}}

\begin{document}

\title{SCAN List Decoding of Polar Codes}

\author{\IEEEauthorblockN{Charles Pillet, Carlo Condo, Valerio Bioglio}\\
\IEEEauthorblockA{Mathematical and Algorithmic Sciences Lab\\ Paris Research Center, Huawei Technologies Co. Ltd.\\
Email: $\{$charles.pillet1,carlo.condo,valerio.bioglio$\}$@huawei.com}} 

\maketitle
\thispagestyle{empty}
\begin{abstract}
In this paper we propose an enhanced soft cancellation (SCAN) decoder for polar codes based on decoding stages permutation.
The proposed soft cancellation list (SCANL) decoder runs $L$ independent SCAN decoders, each one relying on a different permuted factor graph. 
The estimated bits are selected among the $L$ candidates through a dedicated metric provided by the decoders. 
Furthermore, we introduce an early-termination scheme reducing decoding latency without affecting error correction performance.
We investigate the error-correction performance of the proposed scheme under various combinations of number of iterations used, permutation set and early-termination condition.
Simulation results show that the proposed SCANL provides similar results when compared with belief propagation list, while having a smaller complexity. 
Moreover, for large list sizes, SCANL outperforms non-CRC aided successive cancellation list decoding.
\end{abstract}

\section{Introduction}
\label{sec:intro}
Polar codes \cite{ArikanFirst} are drawing increasing interest in both industrial and academic research, especially after their adoption in the 5G wireless standard \cite{3GPP_TS}. 
Polar codes are a class linear block codes relying on channel polarization; they are shown to be capacity-achieving on binary memoryless channels under successive cancellation (SC) decoding for infinite block length. 
List decoding (SCL) was proposed in \cite{TalSCL} to achieve state-of-the-art error-correction performance at finite length, at the cost of higher complexity and latency.
SC and SCL are inherently hard-output decoders, but soft-output decoders have been proposed as well.
Belief Propagation (BP) has been adapted to polar codes in \cite{BPfirst}, however requiring a large number of iterations to achieve SC performance. 
Soft cancellation (SCAN) \cite{SCANfirst} is an iterative decoder that allows to reduce the number of iterations of BP by adopting the SC schedule. 

Permutation decoding has been proposed in \cite{PermGross} to reduce decoding latency with respect to list decoding by running independent SC decoder over equivalent code representations constructed by permuting the stages of the original factor graph.
BP decoding is applied in \cite{BPLCRC} to at most $L$ representations of the polar code, knowing that a polar code of length $N$ has $(\log_2 N)!$ equivalent representations.
A cyclic redundancy check (CRC) code is concatenated to the polar code to stop the decoding. 
Permutation decoding using BP was shown to outperform non CRC-aided SCL block error rate (BLER) performance for a large number of permutations (32). 
Authors in \cite{BPLRM} run the equivalent BP decoders in parallel, proposing the Belief Propagation List (BPL) decoder. 
A parallel SC-based permutation decoding was introduced in \cite{PermDecRussian} showing to approach standard SCL performance with moderate list size, however the discussion is limited to small code lengths. 

In this paper, we propose a new soft-input/soft-output algorithm for polar codes named SCAN List (SCANL) relying on permutation decoding and on parallel SCAN decoders. 
A novel early termination scheme is presented, allowing for latency reduction and energy saving. 
The proposed scheme is easily parallelizable, and approaches non CRC-aided SCL performance with 5 iterations and a list size of 32.

This work is organized as follows. 
In Section~\ref{sec:pre}, polar codes are introduced and the SCAN decoder is detailed. 
In Section~\ref{sec:SCANL}, we describe the proposed SCANL decoder, while in Section~\ref{sec:num} we compare its performance to SC, SCL and known soft-output decoders of polar codes. 
Finally, Section~\ref{sec:conclusions} concludes the paper.

\section{Preliminaries}
\label{sec:pre}
\subsection{Polar Codes}
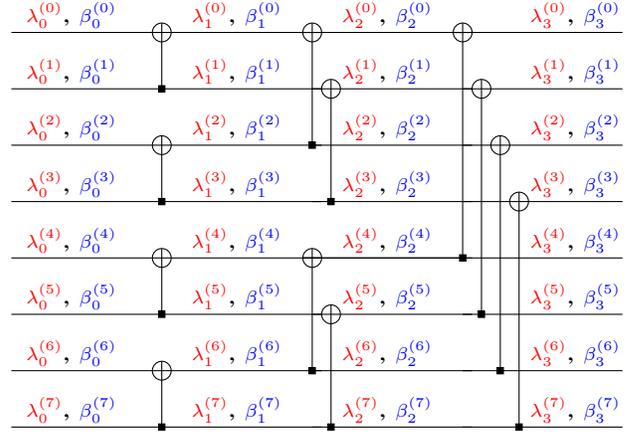
\begin{figure}[t!]
\centering
\begin{tikzpicture}
\newcommand{\polarcodetikz}[7]{
\draw (#1,#2) -- (#5-#6, #2);
\draw (#5,#2) circle(#6cm); 
\draw (#5-#6,#2) -- (#5+#6,#2); 
\draw (#5,#2-#6) -- (#5,#2+#6); 
\draw (#3,#4) -- (#5-#6/3,#4);
\draw [fill=black] (#5-#6/3,#4-#6/3) rectangle (#5+#6/3,#4+#6/3);
\draw (#5,#4+#6/3) -- (#5,#2-#6);
\draw(#5+#6,#2) -- (#5+#6+#7,#2);
\draw(#5+#6/3,#4) -- (#5+#6+#7,#4);
}
\def\Lrec{2.2}
\def\MiddleLrec{\Lrec/2}
\def\Wrec{1.2}
\def\MiddleWrec{\Wrec/2}
\def\Lfirstline{1}
\def\OutputLine{1}
\def\MiddleOutputLine{\OutputLine/2}
\def\MiddleO{0}
\def\UpperO{2}
\def\LowerO{-2}
\def\AbsLeftRec{\Lfirstline}
\def\AbsMiddleRec{\AbsLeftRec+\Lrec+\OutputLine}
\def\AbsRightRec{\AbsMiddleRec+\Lrec+\OutputLine}
\def\AbsRRR{\AbsRightRec+\Lrec+\OutputLine}
\def\Rad{0.125}
\def\gap{0.75}
\def\out{2}
\def\N{7}
\def\logN{3}
\def\begina{0.4*\out}
\def\enda{0.6*\out}
\def\stage{0}
\foreach \i in {0,...,\N}
{

\draw (\begina,-\i*\gap-0.08*\gap) node [above] {\cred{\scriptsize{$\Lm_{\stage}^{(\i)}$}}, \cblue{\scriptsize{$\B_{\stage}^{(\i)}$}}} ;
}
\polarcodetikz{0}{0}{0}{0-\gap}{\out}{\Rad}{\out}
\polarcodetikz{0}{-2*\gap}{0}{-2*\gap-\gap}{\out}{\Rad}{\out}
\polarcodetikz{0}{-4*\gap}{0}{-4*\gap-\gap}{\out}{\Rad}{\out}
\polarcodetikz{0}{-6*\gap}{0}{-6*\gap-\gap}{\out}{\Rad}{\out}
\def\absf{2*\out}
\def\gapf{\gap*2}
\def\begina{\absf-0.4*\out}
\def\enda{\absf-0.2*\out}
\def\stage{1} 
\foreach \i in {0,...,\N}
{

\draw (\absf-0.5*\out,-\i*\gap-0.08*\gap) node [above] {\cred{\scriptsize{$\Lm_{\stage}^{(\i)}$}}, \cblue{\scriptsize{$\B_{\stage}^{(\i)}$}}} ;
}
\polarcodetikz{\absf}{0}{\absf}{0-\gapf}{\absf}{\Rad}{\out}
\polarcodetikz{\absf}{-\gap}{\absf}{-\gap-\gapf}{\absf+0.125*\out}{\Rad}{0.875*\out}
\polarcodetikz{\absf}{-4*\gap}{\absf}{-4*\gap-\gapf}{\absf}{\Rad}{\out}
\polarcodetikz{\absf}{-5*\gap}{\absf}{-5*\gap-\gapf}{\absf+0.125*\out}{\Rad}{0.875*\out}
\def\abst{3*\out}
\def\gapt{\gap*4}

\def\begina{\abst-0.4*\out}
\def\enda{\abst-0.2*\out}
\def\stage{2}
\foreach \i in {0,...,\N}
{
\draw (\abst-0.5*\out,-\i*\gap-0.08*\gap) node [above] {\cred{\scriptsize{$\Lm_{\stage}^{(\i)}$}}, \cblue{\scriptsize{$\B_{\stage}^{(\i)}$}}} ;
}
\polarcodetikz{\abst}{0}{\absf}{0-\gapt}{\abst}{\Rad}{\out}
\polarcodetikz{\abst}{-\gap}{\abst}{-\gap-\gapt}{\abst+0.125*\out}{\Rad}{0.875*\out}
\polarcodetikz{\abst}{-2*\gap}{\abst}{-2*\gap-\gapt}{\abst+0.25*\out}{\Rad}{0.75*\out}
\polarcodetikz{\abst}{-3*\gap}{\abst}{-3*\gap-\gapt}{\abst+0.375*\out}{\Rad}{0.625*\out}
\def\absR{4.25*\out}
\def\begina{\absR-0.4*\out}
\def\enda{\absR-0.2*\out}
\def\stage{3}
\foreach \i in {0,...,\N}
{
\draw (\absR-0.5*\out,-\i*\gap-0.08*\gap) node [above] {\cred{\scriptsize{$\Lm_{\stage}^{(\i)}$}}, \cblue{\scriptsize{$\B_{\stage}^{(\i)}$}}} ;
}
\end{tikzpicture}
\caption{Factor graph of $N=8$ polar codes with soft messages}
\label{fig:scan8}
\end{figure}
A polar code of length $N=2^n$ and dimension $K$ is a block code relying on the polarization effect of kernel matrix $G_2\triangleq \begin{bmatrix}
1 & 0\\
1& 1
\end{bmatrix}$, with transformation matrix $G_N = G_2^{\otimes n}$. 
Polarization creates $N$ virtual bit-channels, each one having a different reliability. 
In an $(N,K)$ polar code, the message bits are stored in the $K$ most reliable bit-channels, that compose the information set $\mathcal{I}$. 
The $N-K$ remaining bit-channels constitute the frozen set $\mathcal{F}$ and are set to a fixed value. 
The computation of bit-channel reliabilities can be performed through Monte Carlo simulation, by tracking the Bhattacharyya parameter or by density evolution under a Gaussian approximation \cite{PolarCodeConstruction}.
In practice, an auxiliary input vector $\mathbf{u}  = \left\{u_0,u_1,\dots, u_{N-1}\right\}$ is generated by assigning $u_i = 0 \text{ if } i \in\mathcal{F}$, and storing information in the remaining entries. 
The codeword $\mathbf{x}$ is then computed as $\mathbf{x} = \mathbf{u}\cdot G_N$.

Successive Cancellation (SC) decoding has been proposed in \cite{ArikanFirst} as the first decoding algorithm for polar codes. 
The decoding process can be described as a binary tree search, where priority is given to the left branch. 
Soft information propagates from the root to the leaves, where bits are estimated, while hard decisions climb the tree towards the root to improve the estimation of the following bits. 
The SCL decoding algorithm proposed in \cite{TalSCL} considers $L$ candidate codewords, improving the error-correction performance of SC at the cost of higher complexity. 
Moreover, a CRC code can be concatenated to polar codes to further improve the performance of SCL \cite{CRCaidedSCL}.

BP decoding is a popular message passing decoder conceived for codes defined on graphs. 
This soft-input/soft-output decoder has been adapted to polar codes in \cite{BPfirst}. 
The factor graph of a polar code can be seen as a graphical representation of it transformation matrix: an example of factor graph is depicted in Figure~\ref{fig:scan8}. 
BP can outperform SC decoding for finite length polar codes after a large number of iterations over the factor graph, implying a high number of operations and thus, energy consumption.

\subsection{SCAN decoding}
\begin{figure}[t!]
	\centering
	\includegraphics[width=1.05\columnwidth]{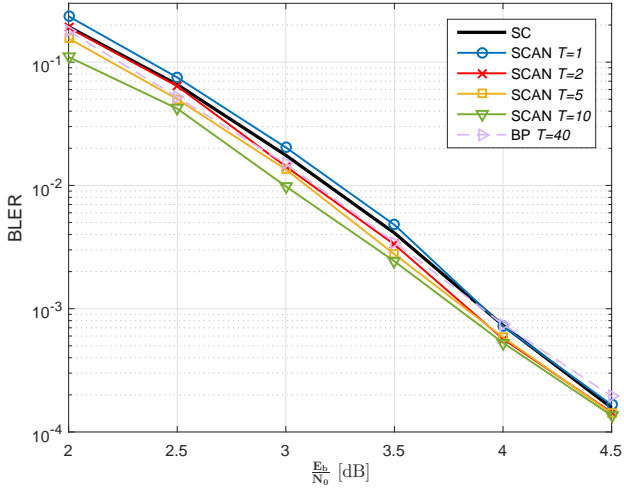}
	\caption{(256,128) BLER with SCAN, SC and BP, for various values of $T$.}
	\label{fig:SCAN}
\end{figure}
The SCAN decoding algorithm has been proposed in \cite{SCANfirst} and is an iterative message passing decoder based on the SC schedule. 
Decoding is performed on the polar factor graph, that can represent both encoding (left to right) and decoding (right to left) (Figure~\ref{fig:scan8}). 
It is composed of $n+1$ stages, each stage having $N$ levels. 
In SCAN decoding, soft information is propagated in both directions: the left-propagating and right-propagating messages at level $0 \leq i < N$ and stage $0\leq s < n+1$ are denoted $\Lm_{s}^{(i)}$ and $\B_{s}^{(i)}$ respectively.
The right hand side left-propagating message $\Lm_{n}^{(i)}$ is initialized according to the received vector $\mathbf{y}$. 
Moreover, the decoder has \emph{a priori} information coming from the frozen set $\mathcal{F}$. 
If the exchanged messages are log-likelihood ratios (LLRs), the message $\B_{0}^{(i)}$ on the left side, relative to the estimated vector $\mathbf{\hat{u}}$, is set to 
\begin{equation}\label{eq:Binit}
\B_{0}^{(i)} =
  \begin{cases}
    \infty \text{,} & \text{if } i \in \mathcal{F}\text{,}\\
    0 \text{,} & \text{otherwise.}
  \end{cases}
\end{equation}
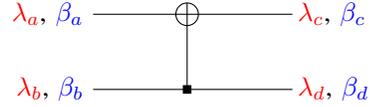
\begin{figure}[t!]
\centering
\begin{tikzpicture}
\newcommand{\polarcodetikz}[7]{
\draw (#1,#2) -- (#5-#6, #2);
\draw (#5,#2) circle(#6cm); 
\draw (#5-#6,#2) -- (#5+#6,#2); 
\draw (#5,#2-#6) -- (#5,#2+#6); 
\draw (#3,#4) -- (#5-#6/3,#4);
\draw [fill=black] (#5-#6/3,#4-#6/3) rectangle (#5+#6/3,#4+#6/3);
\draw (#5,#4+#6/3) -- (#5,#2-#6);
\draw(#5+#6,#2) -- (#5+#6+#7,#2);
\draw(#5+#6/3,#4) -- (#5+#6+#7,#4);
}
\def\Rad{0.15}
\def\gap{1}
\def\out{1.25}
\draw  (0,0) node [left] {\textcolor{red}{$\Lm_a$}, \textcolor{blue}{$\B_a$}};
\draw  (0,-\gap) node [left] {\textcolor{red}{$\Lm_b$}, \textcolor{blue}{$\B_b$}};
\draw  (2*\out+0.1,0) node [right] {\textcolor{red}{$\Lm_c$}, \textcolor{blue}{$\B_c$}};
\draw  (2*\out+0.1,-\gap) node [right] {\textcolor{red}{$\Lm_d$}, \textcolor{blue}{$\B_d$}};
\polarcodetikz{0}{0}{0}{0-\gap}{\out}{\Rad}{\out}
\end{tikzpicture}
\caption{Factor graph with soft messages $\Lm$ and $\B$ for $N = 2$ polar code.}
\label{fig:scan2}
\end{figure}
\begin{figure}[t!]
\centering
\begin{tikzpicture}[scale=1.2]
\def\Lfc{4}
\def\gap{0.5}
\def\radius{0.125}
\def\spacecircle{0.2}
\def\radiusrec{0.06}
\def\firstcircle{0.4}
\def\spacestage{0.4}
\def\spacefc{0.5}
\def\Lfc{2*\firstcircle+7*\spacecircle+2*\spacestage}
\foreach \i in {0,...,7}
{
	\draw (0,-\i*\gap) -- (\Lfc, -\i*\gap);
	\draw (\spacefc+\Lfc,-\i*\gap) -- (\spacefc+\Lfc+\Lfc, -\i*\gap);
}
\foreach \i in {0,...,3}
{
\def\abs{\firstcircle+4*\spacecircle+\spacestage}
\def\ord{-\i*2*\gap}
\def\ordlinked{-\i*2*\gap-\gap}
\draw (\abs,\ord) circle(\radius cm); 
\draw [fill=black] (\abs-\radiusrec,\ordlinked-\radiusrec) rectangle (\abs+\radiusrec,\ordlinked+\radiusrec);
\draw (\abs,\ord+\radius) -- (\abs, \ordlinked);

\def\abs{\Lfc+\firstcircle+\spacefc+2*\spacestage+6*\spacecircle}
\draw (\abs,\ord) circle(\radius cm);
\draw [fill=black] (\abs-\radiusrec,\ordlinked-\radiusrec) rectangle (\abs+\radiusrec,\ordlinked+\radiusrec);
\draw (\abs,\ord+\radius) -- (\abs, \ordlinked);
}
\foreach \i in {0,...,3}
{
	\def\abs{\firstcircle+\i*\spacecircle}
	\def\ord{-\i*\gap}
	\def\ordlinked{\ord-4*\gap}
	\draw (\abs,\ord) circle(\radius cm);
	\draw [fill=black] (\abs-\radiusrec,\ordlinked-\radiusrec) rectangle (\abs+\radiusrec,\ordlinked+\radiusrec);
	\draw (\abs,\ord+\radius) -- (\abs, \ordlinked);
	
	\def\abs{\Lfc+\firstcircle+\spacefc+2*\spacecircle+\spacestage+\i*\spacecircle}
	\draw (\abs,\ord) circle(\radius cm);
	\draw [fill=black] (\abs-\radiusrec,\ordlinked-\radiusrec) rectangle (\abs+\radiusrec,\ordlinked+\radiusrec);
	\draw (\abs,\ord+\radius) -- (\abs, \ordlinked);
}
\foreach \i in {0,...,1}
{
	\def\abs{\firstcircle+2*\spacestage+5*\spacecircle+\i*\spacecircle}
	\def\ord{-\i*\gap}
	\def\ordlinked{\ord-2*\gap}
	
	\draw (\abs,\ord) circle(\radius cm);
	\draw [fill=black] (\abs-\radiusrec,\ordlinked-\radiusrec) rectangle (\abs+\radiusrec,\ordlinked+\radiusrec);
	\draw (\abs,\ord+\radius) -- (\abs, \ordlinked);
	
	\def\ord{-4*\gap-\i*\gap}
	\draw (\abs,\ord) circle(\radius cm);
	\draw [fill=black] (\abs-\radiusrec,\ordlinked-\radiusrec) rectangle (\abs+\radiusrec,\ordlinked+\radiusrec);
	\draw (\abs,\ord+\radius) -- (\abs, \ordlinked);
	
	\def\abs{\Lfc+\spacefc+\firstcircle+\i*\spacecircle}
	\def\ord{-\i*\gap}
	\draw (\abs,\ord) circle(\radius cm);
	\draw [fill=black] (\abs-\radiusrec,\ordlinked-\radiusrec) rectangle (\abs+\radiusrec,\ordlinked+\radiusrec);
	\draw (\abs,\ord+\radius) -- (\abs, \ordlinked);
	\def\ord{-4*\gap-\i*\gap}
	\draw (\abs,\ord) circle(\radius cm);
	\draw [fill=black] (\abs-\radiusrec,\ordlinked-\radiusrec) rectangle (\abs+\radiusrec,\ordlinked+\radiusrec);
	\draw (\abs,\ord+\radius) -- (\abs, \ordlinked);
	
}

\draw [dashed, draw=gray] (\firstcircle+4*\spacecircle+0.5*\spacestage,-7.2*\gap) rectangle (\firstcircle+\spacestage+5*\spacecircle,0.4*\gap); 

\draw [dashed, draw=gray] (\Lfc+\firstcircle+\spacefc+2*\spacestage+6*\spacecircle-0.5*\spacestage,-7.2*\gap) rectangle (\Lfc+\firstcircle+\spacefc+2*\spacestage+7*\spacecircle,0.4*\gap); 

\draw (4.5*\spacecircle+\spacestage+\firstcircle*0.5+0.25*\spacestage,-7.2*\gap) node[below] {$0$};
\draw (\Lfc+\firstcircle+\spacefc+6.5*\spacecircle+1.75*\spacestage,-7.2*\gap) node[below] {$0$};
\draw [dashed, draw=gray] (\firstcircle+2*\spacestage+5*\spacecircle-0.5*\spacestage,-7.2*\gap) rectangle (\firstcircle+2*\spacestage+7*\spacecircle,0.4*\gap); 
\draw [dashed, draw=gray] (\Lfc+\firstcircle+\spacefc-0.5*\spacestage,-7.2*\gap) rectangle (\Lfc+\firstcircle+\spacefc+2*\spacecircle,0.4*\gap); 
\draw (\Lfc+\firstcircle+\spacefc+0.25*\spacestage,-7.2*\gap) node[below] {$1$};
\draw (6*\spacecircle+2*\spacestage+\firstcircle*0.5+0.25*\spacestage,-7.2*\gap) node[below] {$1$};
\draw [dashed, draw=gray] (\firstcircle-0.5*\spacestage,-7.2*\gap) rectangle (\firstcircle+4*\spacecircle,0.4*\gap); 
\draw [dashed, draw=gray] (\Lfc+\firstcircle+\spacefc+2*\spacecircle+\spacestage-0.5*\spacestage,-7.2*\gap) rectangle (\Lfc+\firstcircle+\spacefc+6*\spacecircle+\spacestage,0.4*\gap); 
\draw (2*\spacecircle+\firstcircle*0.5+0.25*\spacestage,-7.2*\gap) node[below] {$2$};
\draw (\Lfc+\spacefc+\spacestage+4*\spacecircle+\firstcircle*0.5+0.25*\spacestage,-7.2*\gap) node[below] {$2$};
\end{tikzpicture}
\caption{Cyclic shift permutations of the factor graph for $N=8$ polar code.}
\label{fig:cyclic}
\end{figure}
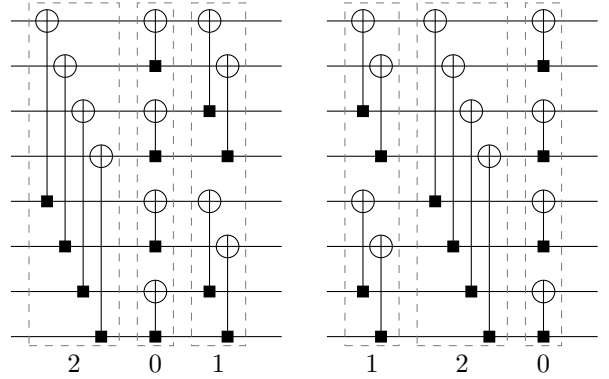
Other messages are initialized to 0 since no further \emph{a priori} information is given. 
It is worth noticing that both sets of messages $\B_{0}$ and $\Lm_{n}$ are not updated through the decoding and keep their initial values. 
Update rules can be described on the basis of the basic polar kernel depicted in Figure~\ref{fig:scan2} as
\begin{align*}
	\cred{\Lm_a} = \tilde{f}\left(\cred{\Lm_c},\cred{\Lm_d} + \cblue{\B_b}\right)\\
	\cred{\Lm_b} = \cred{\Lm_d} + \tilde{f}\left(\cred{\Lm_c}, \cblue{\B_a}\right)\\
	\cblue{\B_c} = \tilde{f}\left(\cblue{\B_a}, \cred{\Lm_d} + \cblue{\B_b}\right)\\
	\cblue{\B_d} = \cblue{\B_b}+\tilde{f}\left(\cblue{\B_a},  \cred{\Lm_c}\right)	
\end{align*}
where $\tilde{f}: \mathbb{R}^2 \rightarrow \mathbb{R}$ is the hardware-friendly implementation of the boxplus operator
\begin{align}\label{eq:fftilde}
	a \boxplus b &\triangleq \log\left(\frac{1+e^{a+b}}{e^a+e^b}\right)\\
	\tilde{f}(a,b) &= \text{min}\left(\left|a\right|,\left|b\right|\right)\text{sign}(a)\text{sign}(b) \simeq a\boxplus b
\end{align}
and decoding follows the SC schedule. 
An iteration of SCAN starts with a left propagation and terminates when $\B_n^{(N-1)}$ has been updated.
After reaching a predefined number of iterations $T$, SCAN decoding stops, taking hard decisions on the left-hand side LLRs as
\begin{equation}
\hat{u}_i =
  \begin{cases}
    0 \text{,} & \text{if } \Lm_{0}^{(i)}+\B_{0}^{(i)} > 0 \text{,}\\ 
    1 \text{,} & \text{otherwise.}
  \end{cases} \label{eq:SCANhard}
\end{equation}
For applications such as product polar codes \cite{PPC}, soft output is required and corresponds to the set $\B_n$.

SCAN converges faster than the BP decoder due to its reliance on the native SC schedule. 
Figure~\ref{fig:SCAN} shows the BLER performance for a (256,128) polar code under SCAN decoding with $T$ iterations; SC and BP with $T=40$ iterations provide a reference. 
It can be seen that two iterations of SCAN are sufficient to outperform both SC and BP decoding. 

\subsection{Permuted factor graph}\label{subsec:pfc}
A polar code of length $N$ has $n!$ equivalent factor graph representations \cite{KoradaPhD} obtained by permuting the $n$ stages of the basic factor graph of the code as shown in Figure~\ref{fig:cyclic}. 
Each representation is defined by a permutation $\pi_i$ of length $n$; among the $n!$ permutations, $n-1$ correspond to a cyclic shift of the trivial permutation $\pi_0$, representing the original stages order. 
Authors in \cite{PermGross} showed that there is a one-to-one mapping between the factor graph permutation and the permutation on codeword positions. Function $\Pi: S_n \rightarrow S_{2^n}$ mapping the factor graph permutation to the codebits permutation, with $S_n=\{0,\dots,n-1\}$, is expressed as 
\begin{align} \label{eq:mapping}
	\Pi\left(\pi_{fc}\right)=\pi(i) = \sum_{s=0}^{n-1}V\left(\pi_{fc}(s),i\right)\cdot2^s ~,
\end{align}
\begin{equation}
V\left(\pi_{fc}(s),i\right) =
  \begin{cases}
    1 \text{,} & \text{if } i \text{ mod}\left(2^{\pi_{fc}(s)+1}\right) \geq 2^{\pi_{fc}(s)}\\ 
    0 \text{,} & \text{otherwise,}
  \end{cases} \label{eq:permSCANhard}
\end{equation}
where $V$ returns 1 if the node on level $i$ and permuted stage $\pi_{fc}(s)$ is a variable node, and 0 otherwise. 
This correlation allows to use the same decoder structure for all the decoding attempts by simply permuting the entries of the incoming LLRs vector \textbf{y}, thus being more implementation-friendly than graph permutation.

The equivalent code representations were exploited in \cite{BPLRM}, that proposes the BP list (BPL) decoder using cyclic shifted factor graphs and $L$ parallel BP decoders. 
BPL outperforms classical BP with the same latency, however without reaching the performance of SCL decoders. But a careful selection of the permutation set may improve decoding performance \cite{PermGross}.
SC-based factor graph permutation decoding was proposed as well to reduce decoding complexity of SCL decoding by eliminating sorting operations \cite{PermDecRussian}. 
This allows to obtain error correction performance comparable to SCL with list size 16 while keeping a decoding latency comparable to SC.

\begin{algorithm}[t!]
 	\def\HiLi{\leavevmode\rlap{\hbox to \hsize{\color{orange!50}\leaders\hrule height .8\baselineskip depth .5ex\hfill}}}
	\SetKwFunction{main}{SCANL}
	\SetKwInOut{Input}{input}%
	\SetKwInOut{Output}{output}%
	\SetKw{Return}{return}
	\SetAlgoLined
	\Input{$N$ channel LLRs $\mathbf{y}$, number of iterations $T$, permutations set $\Pi$, frozen set $\mathcal{F}$}
	\Output{Est. input vector $\mathbf{\hat{u}}$, est. code bits LLRs $\mathbf{s}$}
	\SetKwProg{Fn}{Function}{:}{\Return}
	\Fn{\main{$\mathbf{y}$, $T$, $\Pi$, $\mathcal{F}$}}{
		$\text{PM} \leftarrow \infty$\;
		\For{$l = 0 \dots L-1$}{
			$\text{PM}_l, \mathbf{\hat{u}}_l, \mathbf{s}_l, \leftarrow \texttt{PermSCAN}\left(\mathbf{y}, T, \pi_l, \mathcal{F}\right)$\;
			\If{$\text{PM}_l<\text{PM}$}{
				$\text{PM}\leftarrow\text{PM}_l$\;
				$\mathbf{s}\leftarrow\pi_l^{-1}\left(\mathbf{s}_l\right)$\;
				$\mathbf{\hat{u}}\leftarrow\pi_l^{-1}\left(\mathbf{\hat{u}}_l\right)$\;
			}
		}
		\Return{{\normalfont vector} $\mathbf{\hat{u}}${\normalfont ,} $\mathbf{s}$}
	}
	\caption{SCANL decoder}\label{alg:SCANL}
\end{algorithm}

The choice of the permutation set is crucial to improve the performance of the original decoder. 
Cyclic shifts were initially selected in \cite{KoradaPhD} due to their simplicity and proved to perform better than random permutations.
Authors in \cite{PermGross} propose to recursively populate the set via Monte-Carlo simulation by selecting the $L$ permutations with the highest probability of successful decoding: such a construction requiring long pre-computations, the authors fixed a certain number of stages to limit the search space. 
A permutation set based on Hamming distance (HD) is proposed in \cite{PermDecRussian}, where the elements of the set are selected to maximize Hamming distances among permutations, defined as 
\begin{align} \label{eq:HD}
	\text{HD}\left(\pi_a, \pi_b\right) = n+1 - \sum_{s=0}^{n} \delta_{\pi_a(s),\pi_b(s)}~,
\end{align}
where $\delta$ is the Kronecker delta function. 
Error probability is known for each bit thanks to density evolution, allowing to compute the block error probability for each permutation, since the only change is that the information bits are interleaved. 
The selected permutation set is thus composed of the $L$ permutations with the lowest block error probability that also verify the distance constraint.
This approach results mainly in permutations of the left-hand side graph stages, while the right-hand side remains stable.


\section{SCAN List Decoder}
\label{sec:SCANL}
\begin{algorithm}[t!]
	\def\HiLi{\leavevmode\rlap{\hbox to \hsize{\color{orange!50}\leaders\hrule height .8\baselineskip depth .5ex\hfill}}}
	\def\HiLiEq{\leavevmode\rlap{\hbox to \hsize{\color{orange!50}\leaders\hrule height 1\baselineskip depth .5ex\hfill}}}
	\def\HoEq{\leavevmode\rlap{\hbox to \hsize{\color{red!50}\leaders\hrule height 1\baselineskip depth .5ex\hfill}}}
	\SetKwFunction{Softtohard}{SoftToHard}
	\SetKwFunction{FMain}{PermSCAN}
	\SetKwInOut{Input}{input}%
	\SetKwInOut{Output}{output}%
	\SetKw{Return}{return}
	\SetAlgoLined
	\Input{$N$ channel LLRs $\mathbf{y}$, number of iterations $T$, permutation $\pi$, frozen set $\mathcal{F}$}
	\Output{Metric PM, Estimated input vector $\mathbf{\hat{u}}$, Estimated codebits LLR \textbf{s}}
	\SetKwProg{Pn}{Function}{:}{\Return}	
	\Pn{\FMain{$\mathbf{y}$, $T$, $\pi$, $\mathcal{F}$}}{
		$\text{PM}=0$\;
		$\Lm_n\leftarrow\pi(\mathbf{y})$\;
		\For{$i = 0 \dots N-1$}{		
			\If{$\pi_l(i) \in \mathcal{F}$}{
				$\B_0^{(i)}\leftarrow\infty$\;
			}
			\Else{
				$\B_0^{(i)}\leftarrow0$\;
			}
		}
		\For{$t = 0 \dots T-1$}{
			\For{$i = 0 \dots N-1$}{
				Right-to-left propagation until $\Lm_0^{(i)}$\;
				\If{$\pi_l(i) \in \mathcal{F}$}{
					$\hat{u}_i\leftarrow0$\;
					\HoEq$\text{PM}_l\leftarrow\text{PM}_{l}-\Lm_0^{(i)}$\;\label{alg:PM}	
					\HiLiEq$\text{PM}_l\leftarrow\text{PM}_{l}-\text{min}(\Lm_0^{(i)}, 0)$\;\label{alg:PMet}
				}
				\Else{
					$\hat{u}_i\leftarrow \texttt{LLRsToBinary}(\Lm_{0}^{(i)}+\B_{0}^{(i)})$\;
				}
				Left-to-right propagation with $\B_0^{(i)}$\;
			}
			\HiLi$\mathbf{\hat{x}}_{\pi^{-1}}\leftarrow\pi^{-1}\left(\texttt{LLRsToBinary}\left(\B_n\right)\right)$\;
			\HiLi$\mathbf{\hat{u}}_{\pi^{-1}}\leftarrow\pi^{-1}(\mathbf{\hat{u}})$\;
			\HiLi\If{$\mathbf{\hat{x}}_{\pi^{-1}} == \mathbf{\hat{u}}_{\pi^{-1}}\cdot G_N$}{
				\HiLi \textbf{break}\;
			}
		}
		\Return{\normalfont{PM, }$\mathbf{\hat{u}}${\normalfont ,} $\B_n$}
	}
	\caption{Permuted SCAN decoder}\label{alg:permSCAN}
\end{algorithm}

The proposed SCANL decoder relies on $L$ factor graph representations of a polar code, each one defined by a different permutation belonging to the set $\Pi = \left\{\pi_0,\pi_1,\dots,\pi_{L-1}\right\}$. 
Each of the $L$ representations is decoded independently using a SCAN decoder: the same factor graph is used for all decoders, while the received LLR vector $\mathbf{y}$ is permuted accordingly to $\pi_i$. 
Along with the input vector estimation, each decoder returns a path metric; the codeword with the lowest metric is selected as the output of SCANL decoding, and the corresponding vector $\mathbf{\hat{u}}$ is deinterleaved with the inverse permutation $\pi_l^{-1}$.
The overall SCANL process is detailed in Algorithm \ref{alg:SCANL}, while Algorithm~\ref{alg:permSCAN} describes the permuted SCAN decoder.

The update rule of the path metric is given by
\begin{align}\label{eq:metricSCANL}
	PM = PM - \Lm_0^{(i)} \text{ if } \pi(i) \in \mathcal{F}~.
\end{align}
In contrast with SCL decoding, the path metric is updated only when a frozen bit is encountered (Algorithm~\ref{alg:permSCAN}, line~\ref{alg:PM}); moreover, the metric can also decrease if the sign of the left-propagating message $\Lm_0^{(i)}$ agrees with the existence of a frozen bits, whereas a penalty is applied otherwise. 
SCANL provides less diversity than SCL; it is in fact common that different SCAN decoders provide the same estimated codeword. 
As a consequence, the concatenation of a CRC does not improve heavily the error correction performance of the code as for CRC-aided SCL, since SCL returns $L$ different codewords.


The $L$ permuted SCAN decoders can be run in parallel, since the algorithm does not imply message exchange among processes nor sorting operations between the paths. 
In this case, the latency of SCANL decoder is equal to the latency of a single SCAN decoder with $T$ iterations. 
The number of operations needed is $\mathcal{O}(T L N \log N)$ corresponding to $T$ SCL. 
With the same latency, BPL requires $\frac{N}{\log N}$ more operations.

\subsection{Early-termination} 
\begin{figure}[t!]
	\includegraphics[width=1.05\columnwidth]{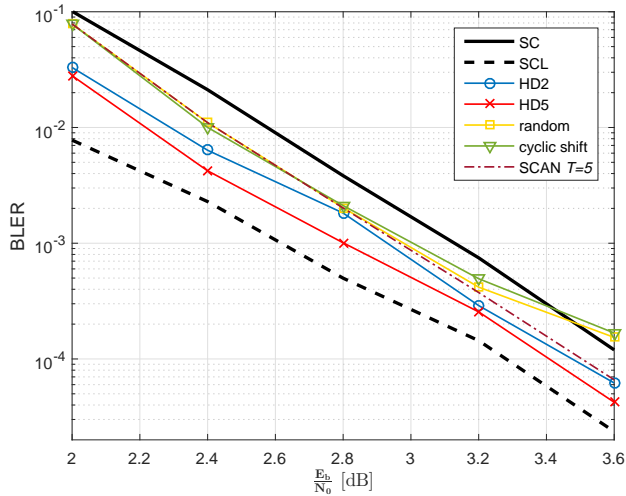}
	\caption{(1024,512) BLER with SCANL, several permutation sets, $L=8$, $T=5$.}
	\label{fig:SCANL}
\end{figure}
We propose a low-complexity early-termination criterion based on the comparison of the hard decision on both left- and right-propagating LLRs. 
This allows to reduce the computational complexity and energy consumption of the decoder by terminating the permuted SCAN decoders after fewer than $T$ iterations. 
If SCAN decoders are run serially, early termination reduces the average latency of SCANL with a negligible error correction performance deterioration. 
In case of parallel implementation, the decoding latency is reduced only when each decoders early-terminates, which is likely to be true for high $E_b/N_0$ and optimized set (Figure~\ref{fig:SCANLET}).

When early termination is activated, the path metric at frozen bits is calculated similarly to SCL, namely without any bonus in case of positive LLRs. 
This avoids giving consecutive bonuses to paths that have not been early-terminated; these would be more likely to have a low $PM$ with respect to terminated paths, that are instead more likely to be correct. 
Early-termination is denoted with the orange lines in Algorithm \ref{alg:permSCAN}, with Line \ref{alg:PMet} substituting Line \ref{alg:PM}. 
Deinterleaving and encoding the estimated input vector after each iterations have a negligible impact on the complexity compared to the decoding on the factor graph. 

\section{Simulation Results}
\label{sec:num}
\begin{figure}[t!]
	\includegraphics[width=1.05\columnwidth]{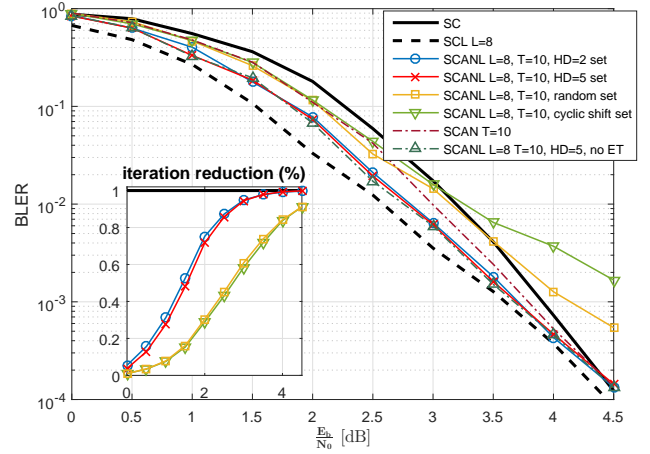}
	\caption{(256, 128) BLER with SCANL and early-termination, $T=10$ and $L=8$.}
	\label{fig:SCANLET}
\end{figure}
Simulation results have been performed over BI-AWGN channel, under BPSK modulation, while the frozen set is designed according to the 5G universal reliability sequence \cite{3GPP_TS}. 
For SCANL decoding, the set of permutations was designed using the Hamming distance method of \cite{PermDecRussian}.
The cyclic shift and random sets have been considered as well for reference. 
Finally, the early-termination mechanism is enabled unless stated otherwise.

Figure~\ref{fig:SCANL} shows the impact of the permutation set on the BLER performance of SCANL decoding, where two Hamming distance (HD) sets are showed with minimum distance 2 and 5. 
We see that the HD-based sets outperform both random and cyclic shift sets. 
Moreover, increasing slightly the distance constraint in the permutation set allows to substantially improve the decoding performance. 

Figure~\ref{fig:SCANLET} shows that early-termination does not deteriorate SCANL performance for HD-based permutation sets, while allowing to reduce decoding latency. 
However, the random and cyclic shift sets are affected by early termination, exhibiting worse performance at high $E_b/N_0$. 
The early termination mechanism substantially reduces the average number of iterations of SCAN; in high $E_b/N_0$ area, the $L$ SCAN decoders early-terminate after 1 iteration instead of the 10 normally required, effectively working with the latency of SC.

\begin{figure}[t!]
	\includegraphics[width=1.06\columnwidth]{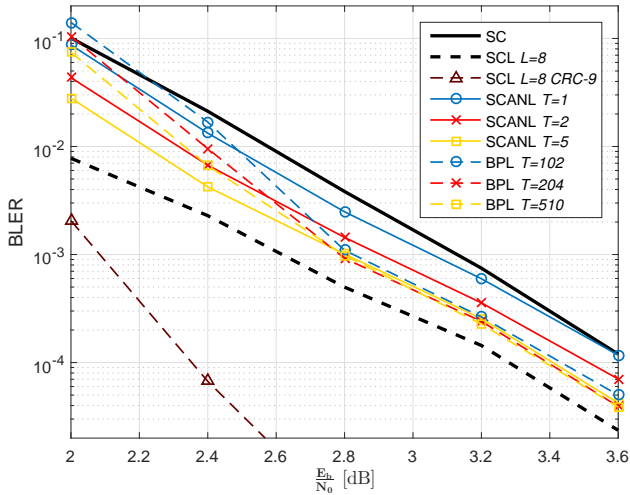}
	\caption{(1024, 512) BLER with SCANL and BPL, $L=8$, when decoded with the same latency.}
	\label{fig:BPL}
\end{figure}

Figure~\ref{fig:BPL} compares the BLER of SCANL with BPL \cite{BPLRM}. 
A list size of $L=8$ is used, whereas the $L$ permutations in the set are selected among the permutations of minimum HD=5 providing the best results for SCANL and BPL.
Both decoders use the same early termination criteria.
Moreover, results are compared at the same latency, allowing BPL to run for $\frac{T_{SCAN}N}{\log_2(N)}$ more iterations ($\approx102\cdot T_{SCAN}$ for $N=1024$), regardless of the higher complexity. 
SCANL outperforms BPL at low $E_b/N_0$, while for high $E_b/N_0$, BPL seems to provide better results. 
BPL does not show significant improvement as $T$ increases, while SCANL matches BPL results at $T=5$ while reducing significantly the number of operations. 
Both decoders show a wide gap from CRC-aided SCL for $L=8$.

\begin{figure}[t!]
	\includegraphics[width=1.05\columnwidth]{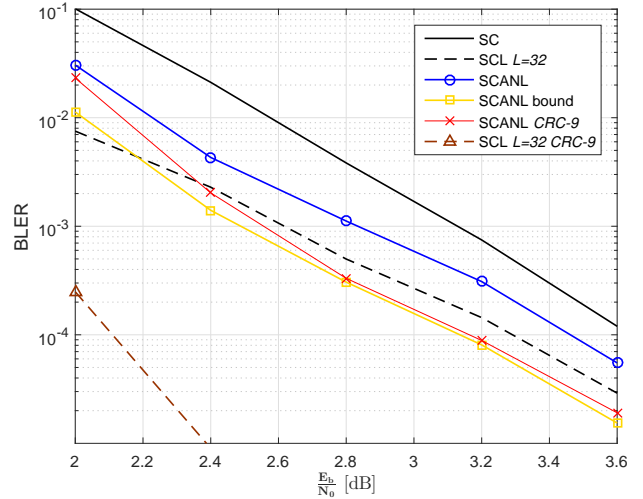}
	\caption{(1024, 512) BLER with SCANL + CRC of length $9$, $L=32$, $T=5$, compared to SCANL, SC, SCL and SCANL bound.}
	\label{fig:crc}
\end{figure}

For larger list sizes, CRC-aided SCANL can beat non-CRC aided SCL, as shown in Figure~\ref{fig:crc}. 
The SCANL bound is obtained by always selecting the candidate for which $\hat{u}=u$, in case it is present. 
It can be seen that CRC-aided SCANL performs close to the SCANL bound, but it is still deeply suboptimal with respect to CRC-aided SCL. 
This is due to the limited candidate diversity of SCANL when compared to SCL, where a higher number of candidates are considered, thanks to path splitting at the information bits. 
The use of a CRC grants a 0.4dB gain for a list size $L=32$.



\section{Conclusions}
\label{sec:conclusions}
In this paper we proposed SCANL, a novel iterative soft-output decoder for polar codes, based on list and permutation decoding. 
This scheme can be run on the same factor graph by simply permuting the received LLRs. 
Moreover, we propose an early-termination scheme allowing to save complexity and latency for a negligible decoding performance cost. 
Compared to SCAN, SCANL exhibits a gain of up to 0.42dB for the same number of inner iterations, when aided with a CRC.
At the same decoding latency, the BLER of SCANL is comparable to that of BPL, with lower decoding complexity. 



\bibliographystyle{IEEEbib}
\bibliography{references}

\end{document}